\documentclass[pre,twocolumn,showpacs]{revtex4}
\usepackage{amsmath}
\usepackage{amssymb}
\usepackage{graphicx}
\usepackage{subfigure}
\newcommand{\dfracd}[2]{\dfrac{\rmd #1}{\rmd #2}}
\newcommand{\average}[1]{\left\langle #1 \right\rangle}
\newcommand{\lessless}{\hspace*{-0.3em}<\hspace*{-0.4em}<\hspace*{-0.3em}}
\newcommand{\rmd}{\text{d}}
\newcommand{\rme}{\text{e}}
\newcommand{\teq}{t_{\text{eq}}}
\newcommand{\tcorr}{t_{\text{corr}}}
\newcommand{\Meq}{M_{\text{eq}}}
\newcommand{\bM}{\boldsymbol{M}}

\begin{document}
\title{Relaxation and Diffusion in a Globally Coupled Hamiltonian System}
\author{YAMAGUCHI Y. Yoshiyuki
  \footnote{
    Electronic address: yyama@i.kyoto-u.ac.jp}}
\affiliation{Department of Applied Mathematics and Physics, 
  Kyoto University, Kyoto, 606-8501, Japan\\
  Dipartimento di Energetica, Universit\`a degli Studi di Firenze,
  Via Santa Marta 3, I-50139 Firenze, Italy}
\date{\today}
\begin{abstract}
  The relation between relaxation and diffusion
  is investigated in a Hamiltonian system 
  of globally coupled rotators.
  Diffusion is anomalous if and only if 
  the system is going towards equilibrium.
  The anomaly in diffusion is not anomalous diffusion
  taking a power-type function,
  but is a transient anomaly due to non-stationarity.
  Contrary to previous claims, in quasi-stationary states,
  diffusion can be explained by a stretched exponential correlation function,
  whose stretching exponent is almost constant
  and correlation time is linear as functions of degrees of freedom.
  The full time evolution is characterized by varying 
  stretching exponent and correlation time.
\end{abstract}
\maketitle

\section{Introduction}
\label{sec:introduction}

Relaxation to thermal equilibrium has been studied
in Hamiltonian systems with long-range interactions
\cite{ruffo-94,antoni-95,tsuchiya-96,yamaguchi-96,torcini-99,latora-02}.
One of the characteristic phenomena in the relaxation process
is anomalous diffusion,
since normal diffusion is expected at equilibrium. 
Anomalous diffusion was firstly investigated 
in a one-dimensional chaotic map
to describe enhanced diffusion in Josephson junctions \cite{geisel-85},
and is observed in many systems both numerically 
\cite{kaneko-89,klafter-94,latora-98,latora-99b,torcini-99}
and experimentally \cite{solomon-93}.

Anomalous diffusion is also observed in
Hamiltonian dynamical systems.
It is explained as due to power-type distribution functions
\cite{klafter-94,meiss-85,zaslavsky-93}
of trapping and untrapping times of the orbit
in the self-similar hierarchy of cylindrical cantori \cite{geisel-95}.
Self-similarity is expected to be one of the important concepts
to understand statistics and motion in Hamiltonian systems,
but cannot be the main feature in systems with many degrees of freedom.
Then, as the first step of approaching the study of self-similarity,
we have to clarify when anomalous diffusion appears,
and what is the origin of the anomaly.

Latora et al. \cite{latora-99b} discussed
the relation between the process of relaxation to equilibrium
and anomalous diffusion in a globally coupled rotator system,
by comparing the time series of the temperature 
and of the mean squared displacement of the phases of the rotators.
They showed that anomalous diffusion changes to a normal diffusion
after a crossover time,
and that the crossover time coincides with the time
when the canonical temperature is reached.
They also claim that anomalous diffusion occurs
in the quasi-stationary states, which appear 
before the system goes towards equilibrium.

The crossover from anomalous to normal diffusion determines the time
when the anomalous diffusion finishes.
However, it is not clearly pointed out when the anomalous diffusion starts,
and hence the study of the relation between the relaxation process
and anomalous diffusion is still not complete.
Moreover, in Ref.\cite{latora-99b}, 
the numerical calculations were performed
by using only one type of initial condition,
but different types of initial condition may change the conclusion
\cite{koyama-01}.

In this article, we study the globally coupled rotator system
considered in Ref.\cite{latora-99b},
and we exhibit the relation between relaxation to equilibrium
and anomalous diffusion with a different type of initial condition
from one used in Ref.\cite{latora-99b}.
Then we show that diffusion becomes anomalous if and only if
the state is neither stationary nor quasi-stationary.
In other words, 
diffusion is shown to be normal in quasi-stationary states,
although a stretched exponential correlation function is present,
contrary to previous claims that report power law type function
\cite{montemurro-02}.
Simple scaling laws of the correlation function imply 
that the result holds irrespective of degrees of freedom.


This paper is organized as follows.
The model, initial condition and observed quantities are
described in Sec.\ref{sec:model}.
In Sec.\ref{sec:relaxation-process}, we study relaxation process,
which we divide into three stages, 
quasi-stationary, relaxational and equilibrium stages.
Diffusion process in each stage is investigated
in Sec.\ref{sec:diffusion-process}
by using stretched exponential correlation functions of momenta.
Dependence on degrees of freedom is also reported
both in Secs.\ref{sec:relaxation-process} and \ref{sec:diffusion-process}.
The last section \ref{sec:summary} is devoted to summary.


\section{Model, Initial Condition and Observed Quantities}
\label{sec:model}

The model considered in this paper has $N$ classical and identical 
rotators confined to move on the unit circle,
and the Hamiltonian is composed of a kinetic and a potential part
\cite{ruffo-94,antoni-95,yamaguchi-96,latora-99b,montemurro-02} ,
\begin{equation}
  \label{eq:model}
  H = K + V = 
  \sum_{j=1}^{N} \dfrac{p_{j}^{2}}{2}
  + \dfrac{1}{2N} \sum_{i,j=1}^{N} [ 1 - \cos (\theta_{i}-\theta_{j}) ].
\end{equation}
The $N$ particles are globally coupled through the mean field
defined as
\begin{equation}
  \label{eq:order-parameter}
  \bM = \dfrac{1}{N} \sum_{j=1}^{N} (\cos\theta_{j}, \sin\theta_{j} )
  = M (\cos\phi, \sin\phi),
\end{equation}
where the modulus $M\ (0\leq M\leq 1)$ represents 
the magnetization of this system.
We remark that the potential $V$ and the kinetic energy $K$
are related to the magnetization $M$ as follows:
\begin{equation}
  \label{eq:V-M}
  2V/N = 1-M^{2},
  \quad
  2K/N = 2U-1+M^{2},
\end{equation}
where $U$ is the energy per particle,
i.e. $U=E/N$, and $E$ is the total energy.
The free energy of this system has been obtained
in the canonical ensemble \cite{ruffo-94,antoni-95,latora-99a},
and it has been shown that
system (\ref{eq:model}) has a second-order phase transition
at the critical energy $U_{c}=0.75$.
If the energy $U$ is greater than the critical energy,
the largest Lyapunov exponent goes to zero
in the thermodynamic limit ($N\to\infty$) \cite{firpo-98}.
Then, all rotators freely rotate,
and diffusion becomes ballistic.
On the contrary, if $U$ is small compared to $U_{c}$,
all rotators are trapped in the potential well
and no diffusion occurs.
We are therefore interested in a value of the energy
which is near but less than the critical energy
in order to allow some particle diffusion.
Hereafter, we set $U=0.69$
(a value studied also in Refs.\cite{latora-99a,latora-99b}).

The canonical equations of motion for system (\ref{eq:model})
can be cast in a form that uses 
the mean field (\ref{eq:order-parameter}) as follows:
\begin{equation}
  \label{eq:eom}
  \dfracd{\theta_{j}}{t} = p_{j}, \quad
  \dfracd{p_{j}}{t} = - M(t) \sin(\theta_{j}-\phi(t)), \quad
  (j=1,\cdots,N).
\end{equation}
We numerically integrate Eq.(\ref{eq:eom})
by using 4th order symplectic integrators \cite{yoshida-93,mclachlan-92}.
The time slice of the integrator is set at $\Delta t=0.2$ or $0.4$,
and it suppresses the relative energy error down to
$| \Delta E/E | < 5 \times 10^{-7}$.

We have performed the integrations starting from $M(0)=0$.
To prepare these initial conditions numerically,
we set $q_{j}(0)=2\pi j/N$, and $p_{j}(0)$
is taken from a uniformly random distribution
whose support is $[-\bar{p},\bar{p}]$, 
where the value $\bar{p}$ is chosen 
to get the energy density $U$.
The total momentum $\sum_{j=1}^{N} p_{j}$ is an integral of the motion
and we initially set it to zero.
This initial state corresponds to a local entropy minimum \cite{antoni-02},
and to a stationary stable solution to the Vlasov-Poisson equation
\cite{antoni-95},
although the system goes towards Gibbs equilibrium
due to finite size effects \cite{latora-02}.
With respect to the $M(0)=1$ initial condition chosen 
in Ref.\cite{latora-99b},
the one we choose has the advantage of being a quasi-stationary state
from the start.

We numerically observe the time series of two quantities.
One is for the relaxation process,
and the other is for the diffusion process. 

To observe the relaxation process, we use the magnetization $M(t)$.
Note that observing $M(t)$ corresponds to observing $2K(t)/N$
by using Eq.(\ref{eq:V-M}), 
and $2K(t)/N$ is the time series of the temperature,
since the canonical average of $2K/N$
coincides with the canonical temperature.

To observe the diffusion process, we introduce
the mean square displacement of phases 
$\sigma_{\theta}^{2}(t)$ defined as
\begin{equation}
  \label{eq:variance}
  \sigma_{\theta}^{2}(t) = \dfrac{1}{N} \sum_{j=1}^{N} 
  [ \theta_{j}(t) - \theta_{j}(0) ]^{2}
  = \average{ [ \theta_{j}(t) - \theta_{j}(0) ]^{2} }_{N}.
\end{equation}
The symbol $\average{\cdot}_{N}$ represents the average 
over all the $N$ rotators.
The quantity $\sigma_{\theta}^{2}(t)$ typically scales as
$\sigma_{\theta}^{2}(t)\sim t^{\alpha}$,
and the diffusion is anomalous when $\alpha\neq 1,2$,
while it is normal when $\alpha=1$ and ballistic for $\alpha=2$.
The quantity $\sigma_{\theta}^{2}(t)$ can be rewritten by using
the correlation function of momenta $C_{p}(t;\tau)$ as
\begin{equation}
  \label{eq:variance2}
  \begin{split}
    \sigma_{\theta}^{2}(t) 
    & = \int_{0}^{t} \rmd t_{1} \int_{0}^{t} \rmd t_{2}~ 
    \average{p_{j}(t_{1})p_{j}(t_{2})}_{N} \\
    & = 2 \int_{0}^{t} \rmd s \int_{0}^{t-s} \rmd \tau~
    C_{p}(s;\tau),
  \end{split}
\end{equation}
where $C_{p}(t;\tau)$ is defined as
\begin{equation}
  \label{eq:corrp}
  C_{p}(t;\tau) = \average{ p_{j}(t+\tau) p_{j}(\tau) }_{N}.
\end{equation}
Moreover, if the system is stationary 
and $C_{p}(t;\tau)$ does not depend on $\tau$ accordingly,
\begin{equation}
  \label{eq:stationarity}
  C_{p}(t;\tau)=C_{p}(t;0), \quad (\forall\tau>0)
\end{equation}
then Eq.(\ref{eq:variance2}) is simplified as
\begin{equation}
  \label{eq:simplified-variance2}
  \sigma_{\theta}^{2}(t) = 2 \int_{0}^{t}
  (t-s)~C_{p}(s;0)~\rmd s.  
\end{equation}


\section{Relaxation Process}
\label{sec:relaxation-process}

\begin{figure}[htbp]
  \centering
  \includegraphics[width=7.5cm]{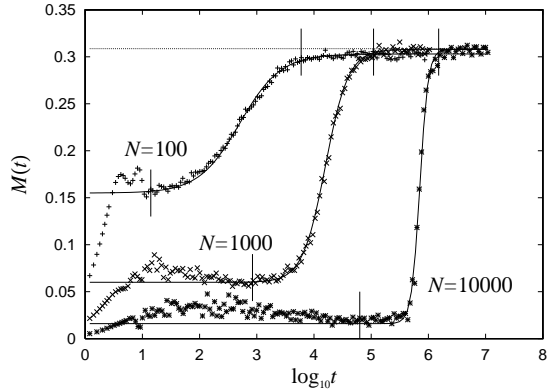}\\
  \caption{Temporal evolutions of $M(t)$.
    $U=0.69$ and $N=100,1000,10000$.
    The horizontal line represents the canonical equilibrium value of $M$.
    On each curve, two short vertical lines are marked.
    The first and the second ones are at the end of Stage-I and II,
    respectively.
    Solid curves are hyperbolic tangent functions
    (\protect\ref{eq:app-tanh-M}). }
  \label{fig:M}
\end{figure}

Temporal evolutions of $M(t)$ are shown in Fig.\ref{fig:M}.
In order to suppress fluctuations,
we have calculated averages over realizations.
Throughout this article, unless no comments appear,
the numbers of realizations are $n=1000,100$ and $8$
for $N=100,~1000$ and $10000$, respectively. 
We divide the temporal evolutions into three stages, 
Stage-I,II and III.
In Stage-I, the value of magnetization is almost constant
but smaller than the canonical value.
After Stage-I, magnetization rapidly increases 
towards its equilibrium value $\Meq$,
and we call this time interval Stage-II.
Finally the system reaches equilibrium, during Stage-III.

\begin{figure}[htbp]
  \centering
  \includegraphics[width=7.5cm]{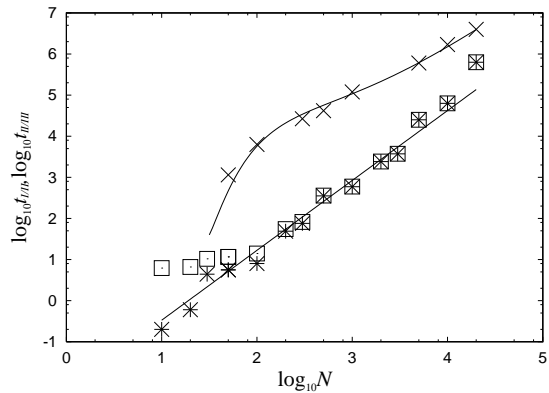}\\
  \caption{Dependence on degrees of freedom of 
    $t_{\text{I/II}}$ (squares) and  $t_{\text{II/III}}$ (crosses).
    Stars represent $t_{\text{I/II}}-6$.
    The lower straight line represents the power law $N^{1.7}/150$.
    The upper curve is a theoretical prediction 
    of the boundary time $t_{\text{II/III}}$ 
    using Eqs.(\protect\ref{eq:parameter-scaling}) and (\protect\ref{eq:t_th})
    with $M_{\text{th}}=0.99\Meq$.}
  \label{fig:scaling}
\end{figure}

\begin{figure}[htbp]
  \centering
  \includegraphics[width=4.1cm]{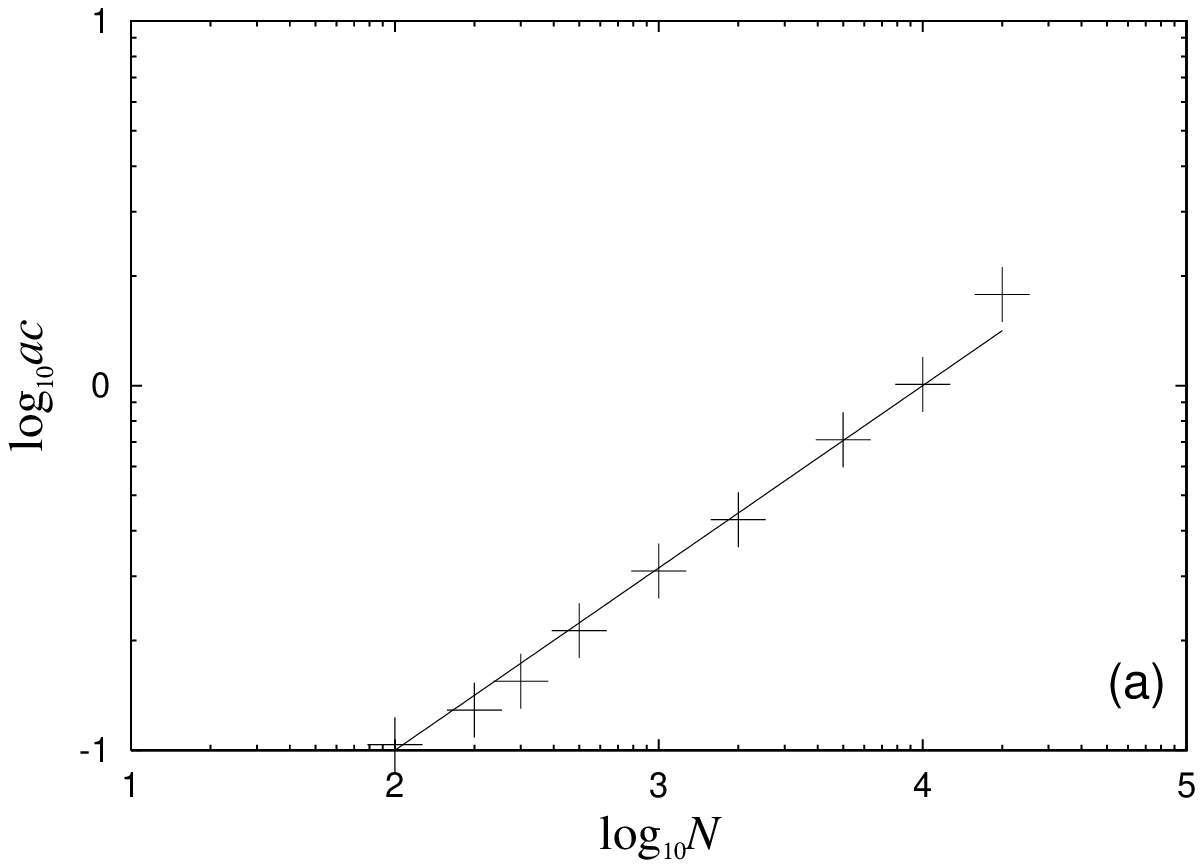}
  \includegraphics[width=4.1cm]{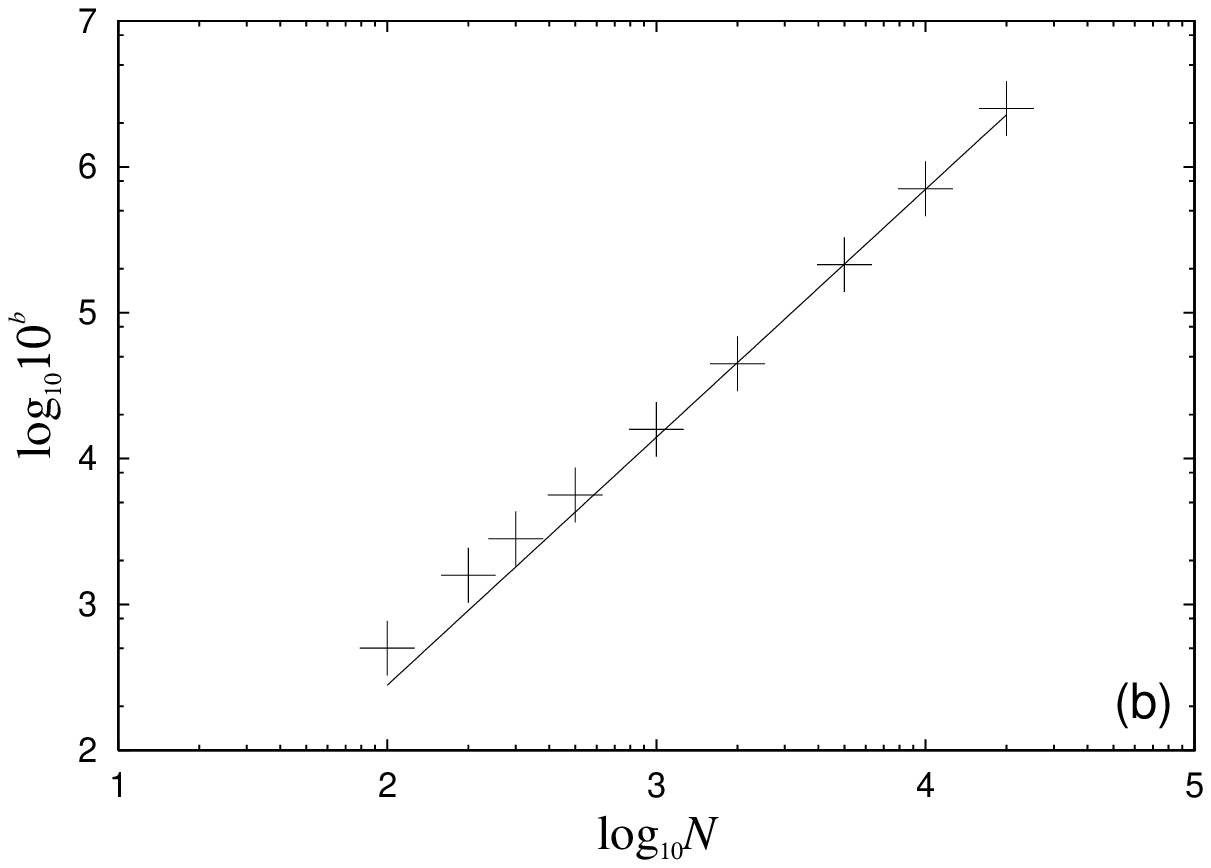}\\
  \includegraphics[width=4.1cm]{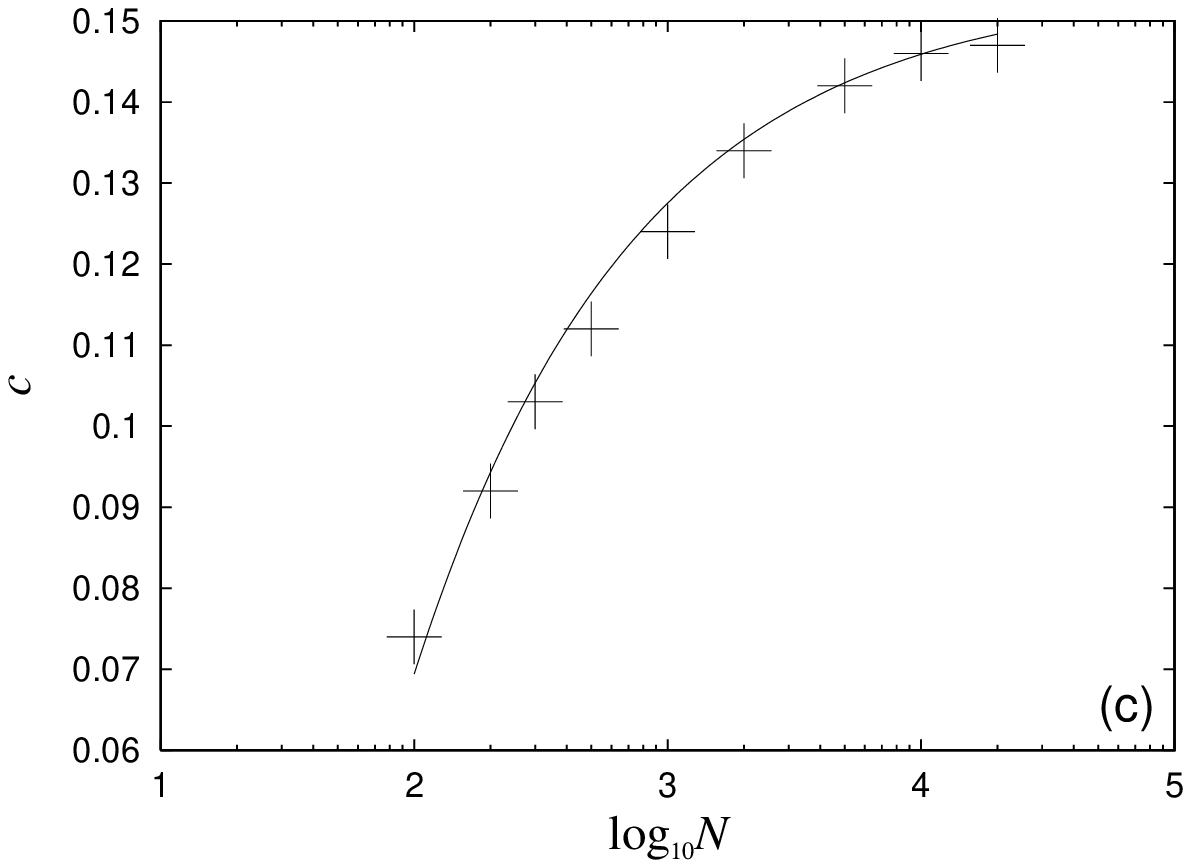}
  \includegraphics[width=4.1cm]{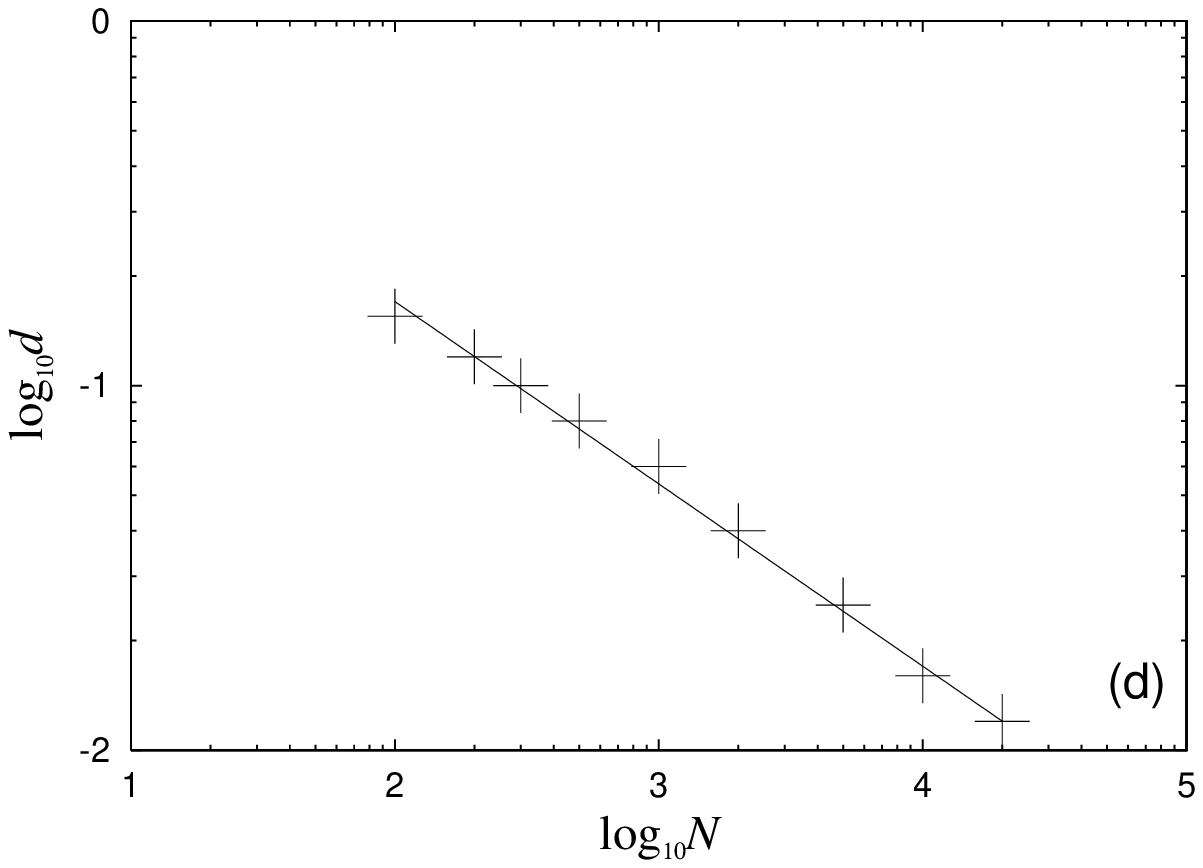}
  \caption{Four parameters $a,~b,c$ and $d$ are reported
    as functions of degrees of freedom.
    (a) Log-log plot of $ac$. (b) Log-log plot of $10^{b}$.
    (c) Liner-log plot of $c$. (d) Log-log plot of $d$.
    Solid curves are scaling functions described 
    in Eq.(\protect\ref{eq:parameter-scaling}).}
  \label{fig:M-scaling}
\end{figure}

Let us define boundary times between Stage-I and II, $t_{\text{I/II}}$,
and between Stage-II and III, $t_{\text{II/III}}$, as follows.
The magnetization takes the local minimum at $t_{\text{min}}$,
and we adopt $t_{\text{I/II}}=t_{\text{min}}$.
We define the other boundary time $t_{\text{II/III}}$ 
as the first passage time which satisfies $M(t)=0.99\Meq$.
Values of the two boundary times are reported in Fig.\ref{fig:scaling}
as functions of degrees of freedom.
The local minimum time is proportional to $N^{1.7}$
for $N\geq 100$ with our initial condition $M(0)=0$,
as with another initial condition $M(0)=1$ \cite{zanette-02}.
For small $N$, we cannot neglect the initial time region $t<6$
in which the level of $M(t)$ goes to $O(1/\sqrt{N})$
coming from the law of large numbers (see Fig.\ref{fig:M-scaling}(d)), 
and hence the power law breaks.
The power law recovers by subtracting the initial increasing time 
$6$ from $t_{\text{I/II}}$ as shown in Fig.\ref{fig:scaling},
i.e. $t_{\text{I/II}}-6 \sim N^{1.7}\ (N\geq 10)$.

A theoretical prediction of $t_{\text{II/III}}$, 
the upper curve in Fig.\ref{fig:scaling},
is obtained by fitting the magnetization $M(t)$ 
as hyperbolic tangent function,
\begin{equation}
  \label{eq:app-tanh-M}
  M(t) = [ 1 + \tanh ( a(\log_{10}t-b )) ]~c + d.
\end{equation}
The parameter $d$ represents the initial level of $M(t)$,
and $c$ the half width between initial and equilibrium levels of $M(t)$.
The product $ac$ is the slope at $\log_{10}t=b$, 
i.e. $ac= \rmd M/\rmd (\log_{10}t)|_{\log_{10}t=b}$,
and $10^{b}$ is the time scale.
As shown in Fig.\ref{fig:M-scaling},
these four parameters are fitted as
\begin{equation}
  \label{eq:parameter-scaling}
  \begin{split}
    & a(N) = \dfrac{\sqrt{N}}{100~c(N)}, \quad
    10^{b(N)} = \dfrac{1}{9} N^{1.7}, \\
    & c(N) = \dfrac{(\Meq-d(N))}{2}, \quad
    d(N) = \dfrac{1.7}{\sqrt{N}}.
  \end{split}
\end{equation}
By using the scaling law (\ref{eq:parameter-scaling}),
we can predict when $M(t)$ reaches a given threshold level,
$M_{\text{th}}$, as a function of $N$.
Let $t_{\text{th}}$ be the threshold time,
which satisfies $M(t_{\text{th}})=M_{\text{th}}$,
then $t_{\text{th}}$ is expressed as
\begin{equation}
  \label{eq:t_th}
  t_{\text{th}} = 10^{b} \left( 
    \dfrac{M_{\text{th}}-d}{\Meq-M_{\text{th}}}
  \right)^{\dfrac{\ln 10}{2a}}.
\end{equation}
In Fig.\ref{fig:scaling}, $t_{\text{th}}$ is reported for 
$M_{\text{th}}=0.99 \Meq$,
and the prediction is in good agreement with numerical results.
We remark that, roughly speaking,
$t_{\text{I/II}}$ is asymptotically proportional to $N^{1.7}$.

The system seems quasi-stationary in Stage-I.
The existence of quasi-stationary states
for sufficiently long time 
has been questioned in Ref.\cite{zanette-02}.
We will answer to the question
by observing dependence on $\tau$ of
the correlation function $C_{p}(t;\tau)$
in Sec.\ref{sec:quasi-stationary}.


\section{Diffusion Process}
\label{sec:diffusion-process}

As described in Eq.(\ref{eq:variance2}),
the mean square displacement $\sigma_{\theta}^{2}(t)$
is obtained from correlation function of momenta $C_{p}(t;\tau)$,
and hence we study diffusion process by observing 
the correlation function.
We start from the simplest stage, Stage-III,
because we may use the simple expression (\ref{eq:simplified-variance2}).
Next, we progress to Stage-I, 
where we expect that the system is quasi-stationary,
and that we may use Eq.(\ref{eq:simplified-variance2}) again.
In non-stationary stage, Stage-II,
we check whether diffusion is of a power type.
Finally we investigate dependence on degrees of freedom
for some important parameters.

\subsection{Diffusion at Equilibrium}
\label{sec:equilibrium}

Assuming the system has reached equilibrium at $t=\teq$,
we observe $C_{p}(t;\teq)$ and $\sigma_{\theta}^{2}(t;\teq)$, where 
\begin{equation}
  \label{eq:time-shift}
  \sigma_{\theta}^{2}(t;\teq) 
  = \average{ [\theta_{j}(t+\teq)-\theta_{j}(\teq)]^{2} }_{N} .
\end{equation}
At equilibrium we may assume that the system is stationary, 
\begin{equation}
  \label{eq:stationarity-equili}
  C_{p}(t;\teq+\tau)=C_{p}(t;\teq), \quad (\forall\tau>0)
\end{equation}
and hence
\begin{equation}
  \label{eq:simplified-variance2-equili}
  \sigma_{\theta}^{2}(t;\teq) = 2 \int_{0}^{t}
  (t-s)~C_{p}(s;\teq)~\rmd s.  
\end{equation}

\begin{figure}[htbp]
  \centering
  \includegraphics[width=7.2cm]{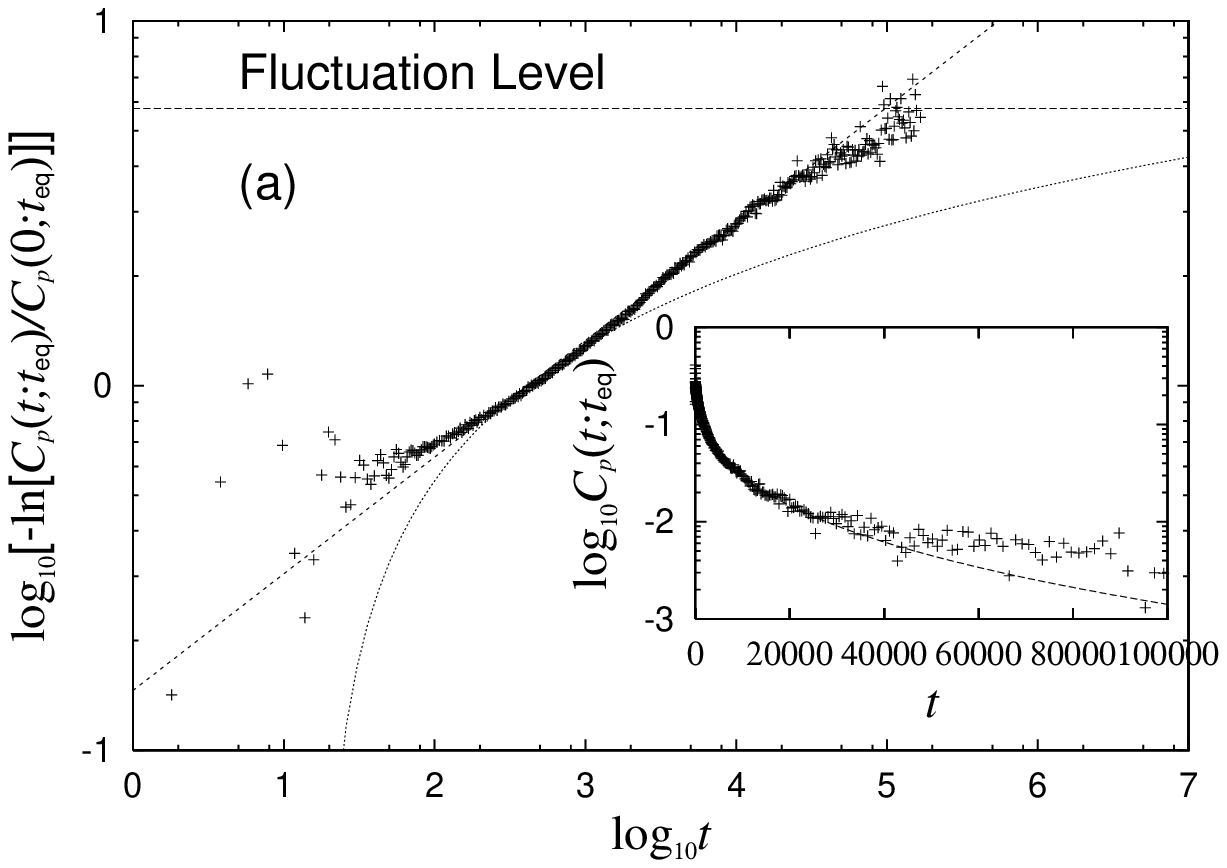}
  \includegraphics[width=7.2cm]{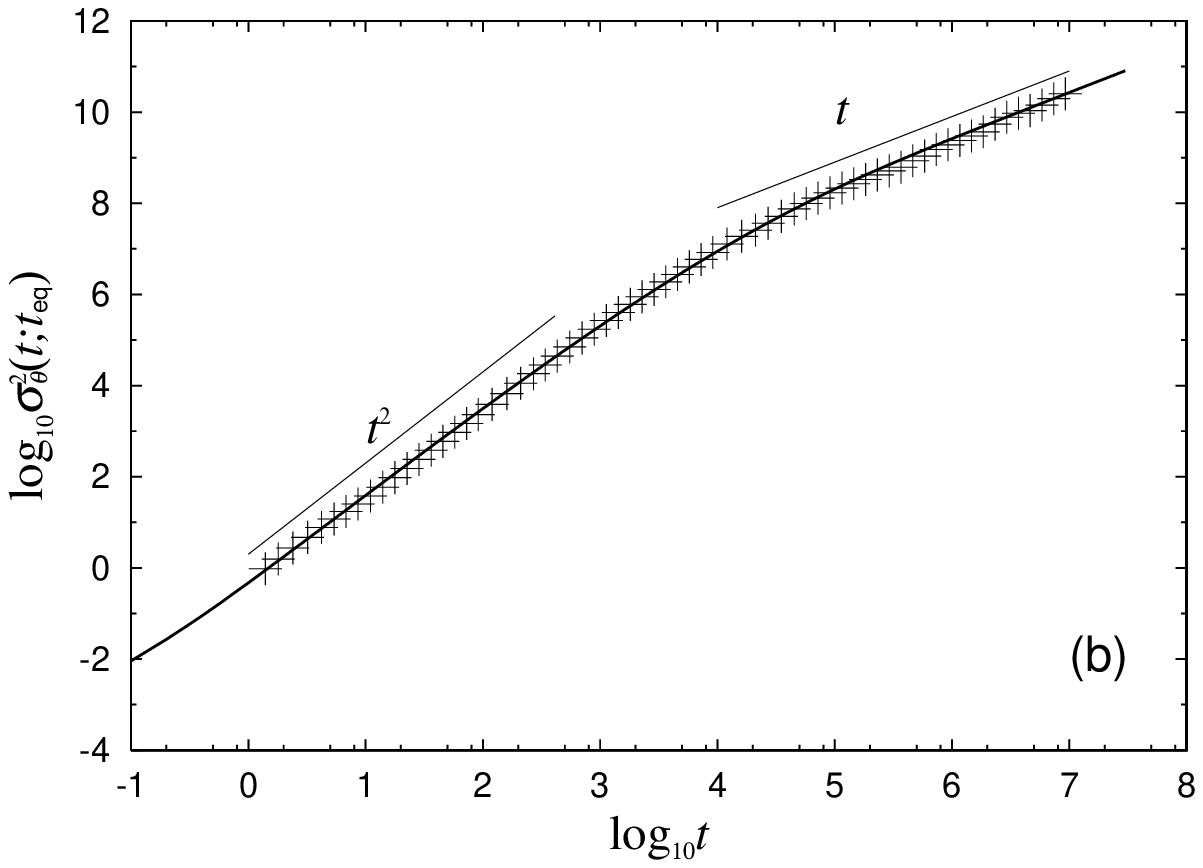}
  \caption{
    (a) Double log-log plot of 
    normalized correlation function $C_{p}(t;\teq)/C_{p}(0;\teq)$ 
    at equilibrium with $\teq=2^{20}$. $N=1000$. 
    We take an average over $n=100$ realizations.
    The straight line and the curve represent 
    the stretched exponential function (\protect\ref{eq:corrp-equili})
    and the power-type function $(t/410)^{-0.32}/\rme$, respectively.
    The upper horizontal line is fluctuation level 
    $O(1/\sqrt{Nn})$.
    The inset shows a log-linear plot with the stretched exponential function.
    (b) Log-log plot of $\sigma_{\theta}^{2}(t;\teq)$
    with the approximate function produced by
    Eqs.(\protect\ref{eq:simplified-variance2-equili})
    and (\protect\ref{eq:corrp-equili}).}
  \label{fig:equili}
\end{figure}

Now let us consider the correlation function for $N=1000$.
We adopt $\teq=2^{20} \simeq 10^{6}$ 
which is long enough to reach equilibrium imaging from Fig.\ref{fig:M}.
The correlation function $C_{p}(t;\teq)$ is reported 
in Fig.\ref{fig:equili}(a),
and is well approximated by the stretched exponential function 
\cite{phillips-96}
\begin{equation}
  \label{eq:corrp-equili}
  C_{p}(t;\teq) = 0.47 \exp[- (t/410)^{0.32} ],
\end{equation}
rather than by a pure exponential
(see the inset of Fig.\ref{fig:equili}(a)
which is a log-linear plot of $C_{p}(t;\teq)$).

We remark that a stretched exponential function $\exp[-x^{\beta}]$
with a small exponent $|\beta|\lessless 1$ 
is indistinguishable from a power-type function 
in the region $|\beta\ln x| \lessless 1$: 
\begin{displaymath}
  \exp[- x^{\beta}] = \exp[ - \exp (\beta\ln x) ]
  \sim \exp[ - 1 - \beta\ln x ] = x^{-\beta} / \rme.
\end{displaymath}
However the fitting function (\ref{eq:corrp-equili}) well agree with
numerical result even around $|0.32\ln (t/410)|=1$,
whose two solutions are $t\simeq 18,~9330$.
We therefore adopt a stretched exponential function
as an approximation of $C_{p}(t;\teq)$.

By using the fitting function (\ref{eq:corrp-equili}) and
Eq.(\ref{eq:simplified-variance2-equili}),
we numerically reproduce $\sigma_{\theta}^{2}(t;\teq)$,
and the reproduced curve well approximates the numerical result
as shown in Fig.\ref{fig:equili}(b).
Note that $\sigma_{\theta}^{2}(t;\teq)$ is proportional to $t^{2}$
in the limit of $t\to 0$, since $C_{p}(s;\teq)$ 
in Eq.(\ref{eq:simplified-variance2-equili}) 
goes to the constant $C_{p}(0;\teq)$.
On the other hand, in the limit of $t\to\infty$,
$\sigma_{\theta}^{2}(t;\teq)$ is proportional to $t$,
because both $C_{p}(s;\teq)$ and $s C_{p}(s;\teq)$ 
are almost zeros in long time region,
and hence their integrals become constants.
The crossover from $t^{2}$ to $t$ is also observed
if we assume an exponential correlation function,
and hence we conclude that diffusion at equilibrium is normal as expected
although a stretched exponential is present.


\subsection{Diffusion in Quasi-Stationary State}
\label{sec:quasi-stationary}

Except for Stage-III, we cannot expect stationarity to hold
(\ref{eq:stationarity}) any more.
However, from the temporal evolutions of $M(t)$, Fig.\ref{fig:M},
we may expect quasi-stationarity in Stage-I,
\begin{equation}
  \label{eq:quasi-stationarity}
  C_{p}(t;\tau) = C_{p}(t;0) + \epsilon(t;\tau),
\end{equation}
where $\tau$ belongs to Stage-I
and $\epsilon(t)$ is suitably small. 

\begin{figure}[htbp]
  \centering
  \includegraphics[width=7.5cm]{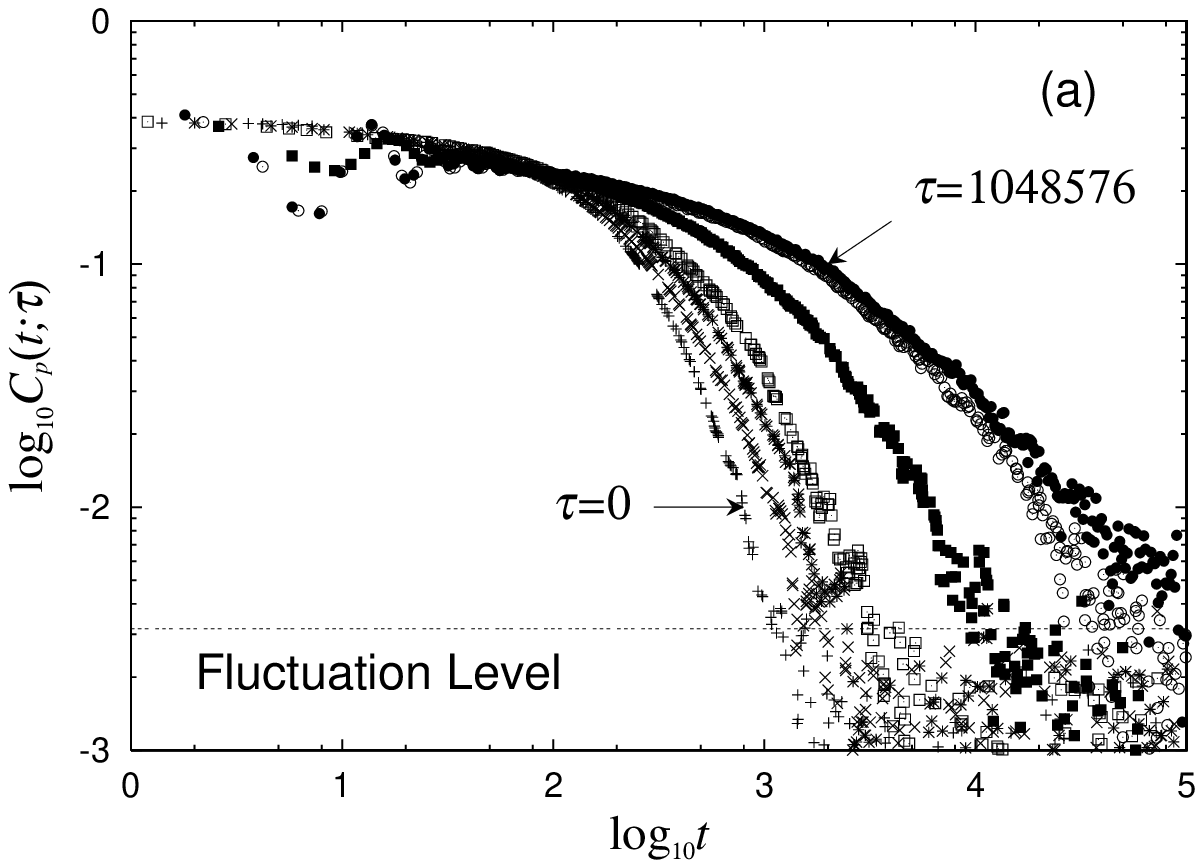}
  \includegraphics[width=7.5cm]{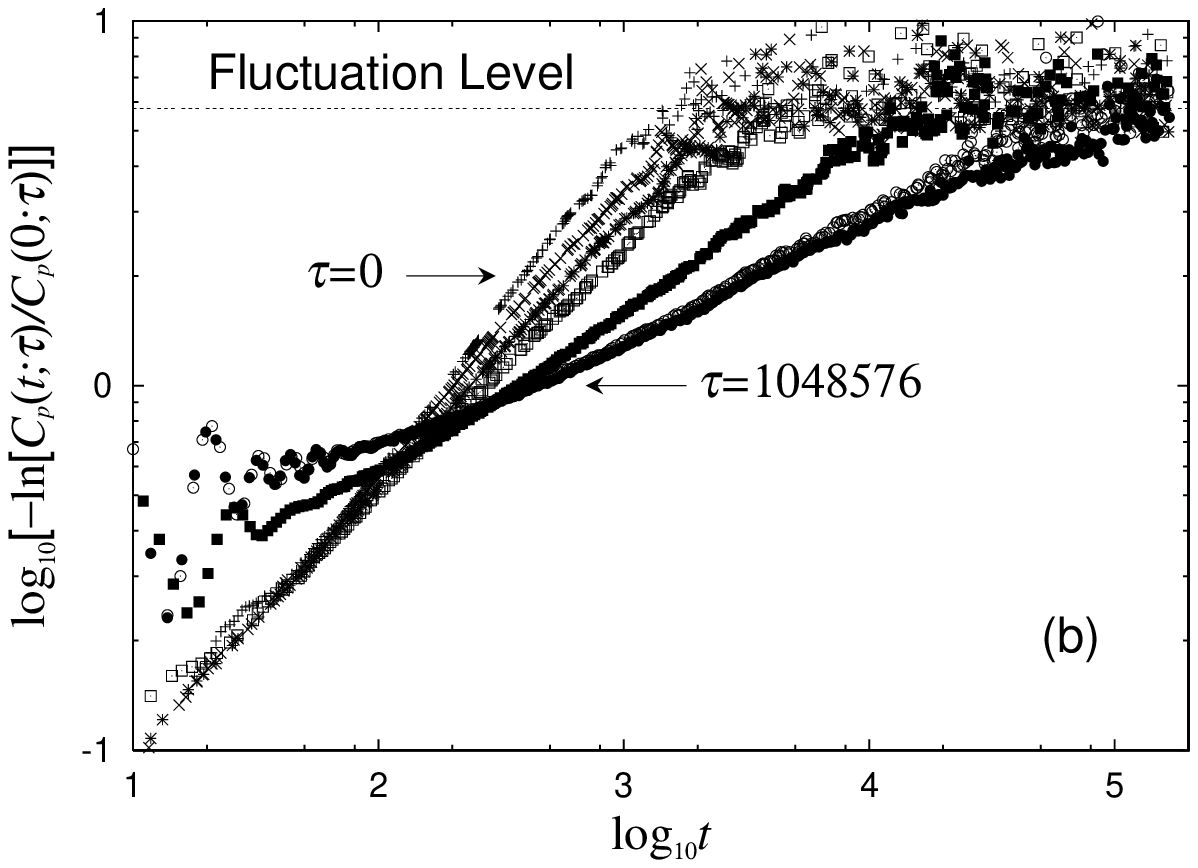}
  \includegraphics[width=7.5cm]{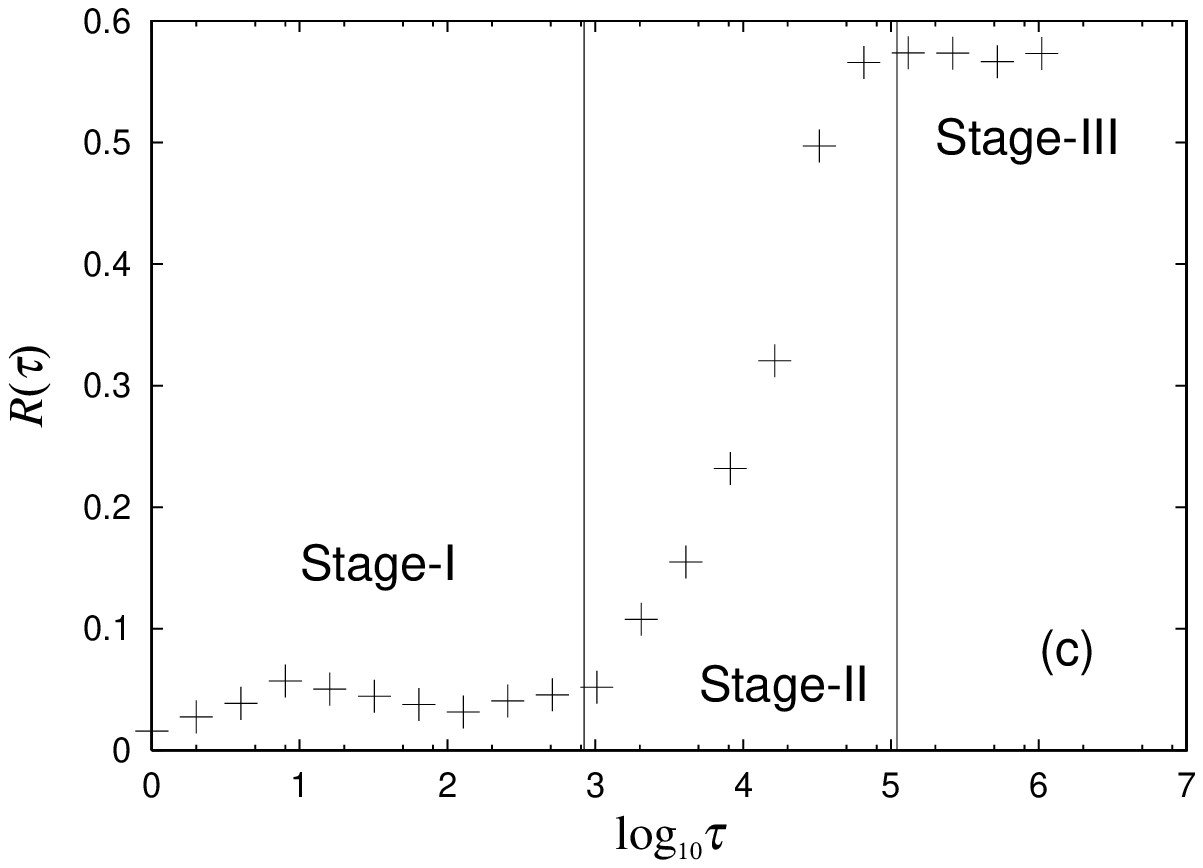}
  \caption{
    Correlation function of momenta $C_{p}(t;\tau)$
    for various values of $\tau=0,1024,2048,4096,16384,65536$
    and $1048576$ from left to right.
    (a) Log-log plot. (b) Double log-log plot.
    (c) Relative error (\ref{eq:relative-error-Cp}) of $C_{p}(t;\tau)$
    from $C_{p}(t;0)$.}
  \label{fig:cpttau}
\end{figure}

The correlation function $C_{p}(t;\tau)$ for various values of $\tau$
is reported in Figs.\ref{fig:cpttau}(a) and (b) for $N=1000$,
and the relative error of correlation function, defined as
\begin{equation}
  \label{eq:relative-error-Cp}
  R(\tau) = \max_{t} \dfrac{ | \epsilon(t;\tau) | }{C_{p}(0;0)},
\end{equation}
is also reported in Fig.\ref{fig:cpttau}(c) as a function of $\tau$.
The error $R(\tau)$ stays small up to the end of Stage-I,
and hence we conclude that the system is quasi-stationary in Stage-I.
We believe that the quasi-stationary states correspond 
to stationary stable states of the Vlasov equation \cite{yamaguchi-03}.
We remark that $R(\tau)$ is constant in Stage-III again
due to stationarity at equilibrium.

It seems natural that we regard $C_{p}(t;\tau)$ 
as a series of stretched exponential functions of $t$
rather than power-type functions,
since this function fits $C_{p}(t;\tau)$
in more than two decades of time
(power law fits of the correlation functions hold in one decade).
Moreover, at equilibrium,
$C_{p}(t;\teq)$ is also a stretched exponential 
rather than a pure exponential,
as shown in Fig.\ref{fig:equili}.

\begin{figure}[htbp]
  \includegraphics[width=7.5cm]{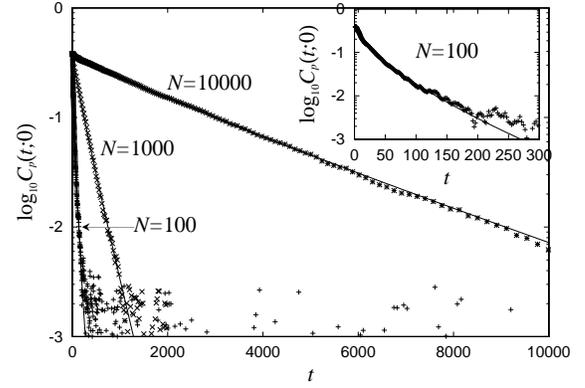}
  \caption{Correlation function of momenta at $\tau=0$, i.e. $C_{p}(t;0)$.
    The inset is magnification of the horizontal axis around $t=0$
    for $N=100$.
    These numerical results are approximated by solid curves
    which are stretched exponential functions (\ref{eq:Cp0}).}
  \label{fig:corrp}
\end{figure}

\begin{figure}[htbp]
  \centering
  \includegraphics[width=7.5cm]{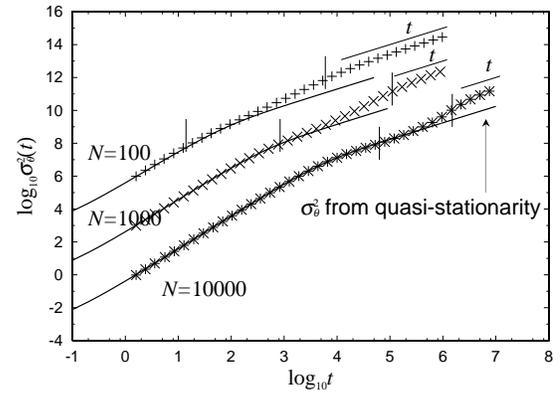}
  \caption{
    Time series of the mean square displacement of the phases 
    $\sigma_{\theta}^{2}(t)$. 
    $N=100,~1000$ and $10000$ from top to bottom.
    The vertical axis is the original scale only for $N=10000$,
    and is multiplied by $10^{3}$ and $10^{6}$ 
    for $N=1000$ and $100$ respectively just for a graphical reason.
    In Stage-I where the system is quasi-stationary,
    the numerical results are approximated by solid curves
    which are obtained from Eq.(\protect\ref{eq:simplified-variance2})
    using functions (\ref{eq:Cp0}).
    After the system reaches equilibrium,
    diffusion becomes normal.
    Anomaly in diffusion is observed only in Stage-II.
    The two short vertical lines on each curve show
    the end of Stages-I and II,
    which correspond to the ones found in Fig.\protect\ref{fig:M}.}
  \label{fig:sigma}
\end{figure}

In the quasi-stationary region, Stage-I,
the mean square displacement $\sigma_{\theta}^{2}(t)$
can be derived by the correlation function $C_{p}(t;0)$,
which is reported in Fig.\ref{fig:corrp} for $N=100,1000$ and $10000$.
We approximate $C_{p}(t;0)$ by a stretched exponential function as
\begin{equation}
  \label{eq:Cp0}
  \begin{split}
    N=100 : \quad & C_{p}(t;0) = 0.38~\exp[ - (t/20)^{0.68} ], \\
    N=1000 : \quad & C_{p}(t;0) = 0.38~\exp[ - (t/180)^{0.91} ], \\
    N=10000 : \quad & C_{p}(t;0) = 0.38~\exp[ - (t/2200)^{0.90} ].
  \end{split}
\end{equation}
The prefactor $0.38$ comes from $C_{p}(0;0)=2K(0)/N$.

Using the approximate functions (\ref{eq:Cp0})
and Eq.(\ref{eq:simplified-variance2}),
we are able to reproduce $\sigma_{\theta}^{2}(t)$,
as shown in Fig.\ref{fig:sigma}.
The approximation is good in Stage-I, 
i.e. in the quasi-stationary time region,
irrespective of the value of $N$.
Consequently, there is no anomaly in diffusion in Stage-I,
since the diffusion is explained 
by stretched exponential correlation function.

\subsection{Diffusion in Non-Stationary State}
\label{sec:non-stationary}

\begin{figure}[htbp]
  \centering
  \includegraphics[width=7.5cm]{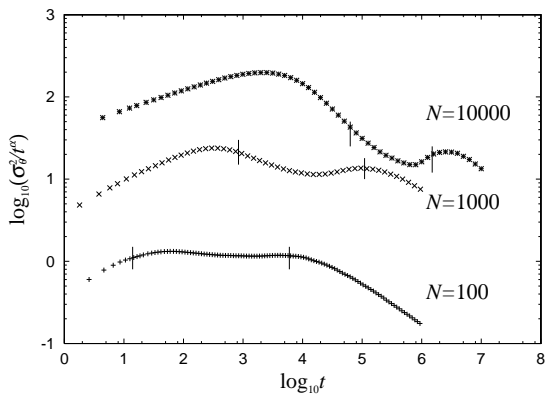}
  \caption{Log-log plot of $\sigma_{\theta}^{2}(t)/t^{\alpha}$.
    The exponent $\alpha$ is estimated as 
    $1.54,~1.59$ and $1.74$ for $N=100,~1000$ and $10000$ respectively.
    The two short vertical lines on each curve show
    the end of Stages-I and II.
    In Stage-II, $\sigma_{\theta}^{2}/t^{\alpha}$ is not constant.
    The vertical axis is the original scale only for $N=100$,
    and is multipled by $10$ and $100$ for $N=1000$ and $N=10000$ 
    respectively for a graphical reason.}
  \label{fig:sigma-divided-talpha}
\end{figure}

After the quasi-stationary region,
diffusion becomes anomalous, which is faster than normal diffusion,
in Stage-II.
If we fit $\sigma_{\theta}^{2}(t)$ by a power-type function $t^{\alpha}$
in Stage-II,
the exponent $\alpha$ is estimated as 
$1.54,~1.59$ and $1.74$ for $N=100,~1000$ and $10000$ respectively.
The values of exponent tend to increase as $N$ increases
as reported for the system having the so-called 
$2$-dimensional egg-crate potential \cite{torcini-99}.
On the other hand, the duration in which diffusion is anomalous
becomes shorter and shorter in logarithmic time scale as $N$ increases , 
in accordance with the sharper change of $M(t)$.
Moreover, $\sigma_{\theta}^{2}/t^{\alpha}$ is not constant,
but has a wave in Stage-II (see Fig.\ref{fig:sigma-divided-talpha}).
Hence we guess that the anomaly in diffusion is not anomalous diffusion
taking a power-type function,
but a transient anomaly due to non-stationarity of Stage-II.

Let us proceed to investigate the origin of anomaly in diffusion.
We focus on the behavior for $N=1000$.
The mean square displacement $\sigma_{\theta}^{2}(t)$ 
is perfectly determined by the correlation function 
$C_{p}(t;\tau)$ using Eq.(\ref{eq:variance2}),
once we assume that $C_{p}(t;\tau)$ is a series of
stretched exponential functions.
We introduce three parameters, $C_{p}(0;\tau)$, $\tcorr(\tau)$ and 
$\beta(\tau)$, to describe the stretched exponential function as
\begin{equation}
  \label{eq:stexp}
  C_{p}(t;\tau) = C_{p}(0;\tau) 
  \exp [ - \{t/\tcorr(\tau)\}^{\beta(\tau)} ].
\end{equation}
We investigate which of the three parameters is the most important
to yield anomaly in diffusion.

The strategy is as follows.
We reproduce $\rmd\sigma_{\theta}^{2}(t)/\rmd t$
by using the three parameters and the formula
\begin{equation}
  \label{eq:dsigma-dt}
  \dfracd{\sigma_{\theta}^{2}}{t}(t)
  = 2 \int_{0}^{t} \rmd \tau~ C_{p}(0;\tau) 
  \exp [ - \{ (t-\tau)/\tcorr(\tau) \}^{\beta(\tau)} ].
\end{equation}
We consider the first derivative of $\sigma_{\theta}^{2}$
instead of $\sigma_{\theta}^{2}$ itself,
because the former requires only single integration
while the latter requires double integrations (\ref{eq:variance2}).
We first omit the dependence on $\tau$ of the parameter $C_{p}(0;\tau)$
and fix it to a constant value to observe 
how it affects the anomaly in diffusion.
We then fix the two other parameters
$\tcorr(\tau)$ and $\beta(\tau)$
to determine their effect on the mean square displacement.

From the numerical results of $C_{p}(t;\tau)$, Fig.\ref{fig:cpttau}(b),
we determine the values of three parameters
$C_{p}(0;\tau),~\tcorr(\tau)$ and $\beta(\tau)$ 
at some value of $\tau$ by using the least square method.
The discrete values of the parameters are not enough
to reproduce $d\sigma_{\theta}^{2}(t)/dt$ accurately,
and then we approximate the parameters
by hyperbolic tangent functions as follows:
\begin{equation}
  \label{eq:app-tanh}
  \begin{split}
    C_{p}(0;\tau) & = 0.046~[1+\tanh (2.5(\log_{10}\tau-4.35)) ] + 0.385,\\
    \tcorr(\tau) & = 80~[1+\tanh (1.5(\log_{10}\tau-3.4))] + 170, \\
    \beta(\tau) & = 0.31~[1+\tanh (1.5(\log_{10}\tau-3.8))] + 0.29. 
  \end{split}
\end{equation}

\begin{figure}[htbp]
  \centering
  \includegraphics[width=7.5cm]{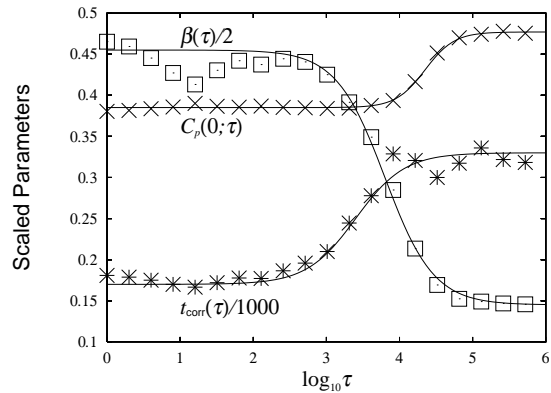}
  \caption{
    The three parameters $C_{p}(0;\tau),~\tcorr(\tau)$ and $\beta(\tau)$
    as functions of $\tau$.
    The latter two parameters $\tcorr(\tau)$ and $\beta(\tau)$ 
    are multiplied by $1/1000$ and $1/2$ respectively 
    for a graphical reason.
    Solid curves are hyperbolic tangent functions
    described in Eq.(\protect\ref{eq:app-tanh}).}
  \label{fig:stexp-factor}
\end{figure}

\begin{figure}[htbp]
  \centering
  \includegraphics[width=4.25cm]{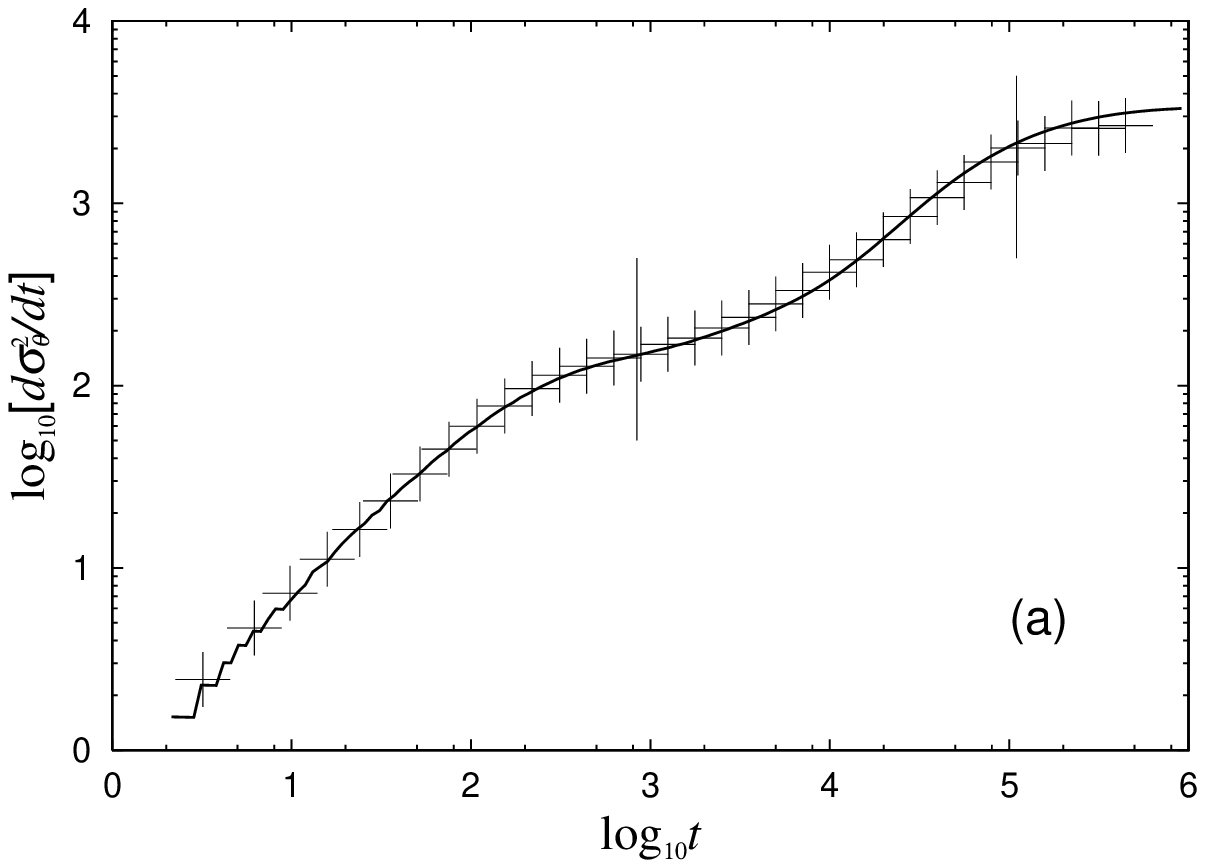}
  \includegraphics[width=4.25cm]{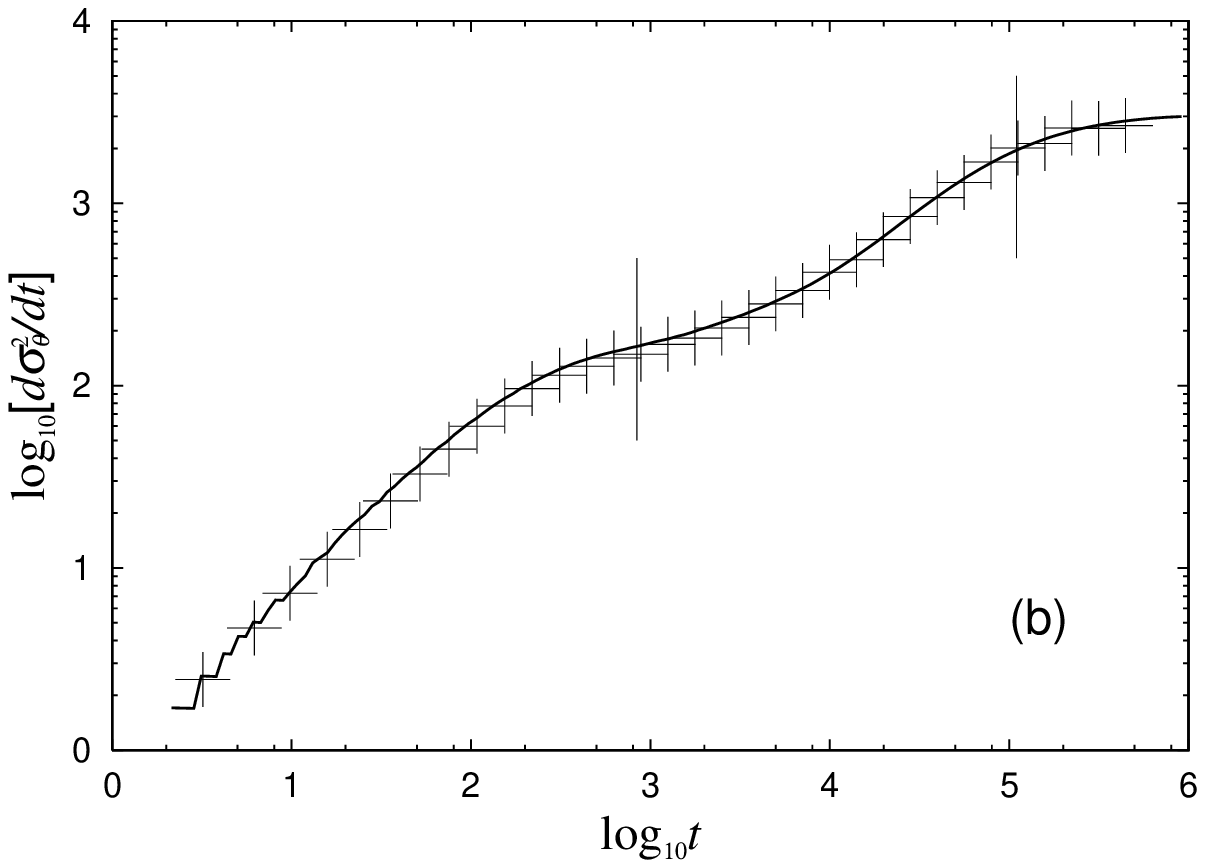}
  \includegraphics[width=4.25cm]{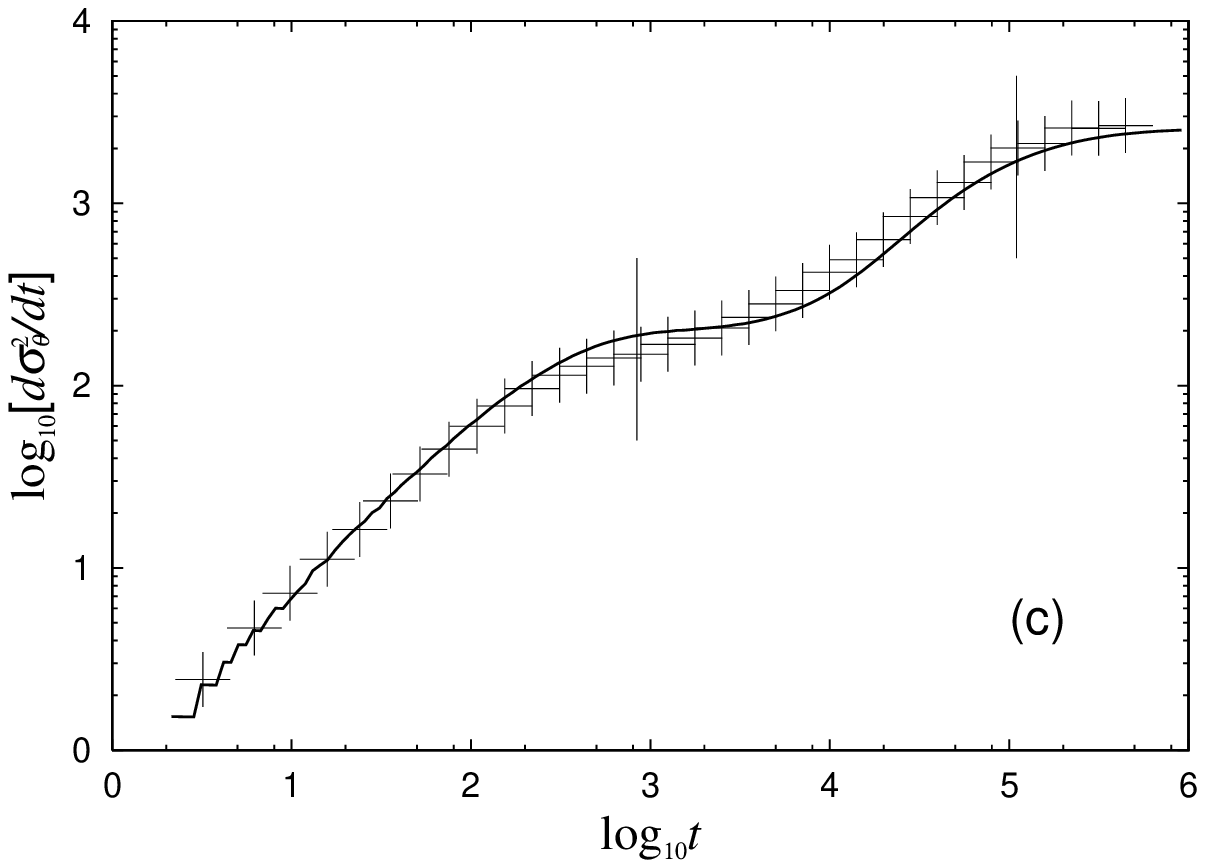}
  \includegraphics[width=4.25cm]{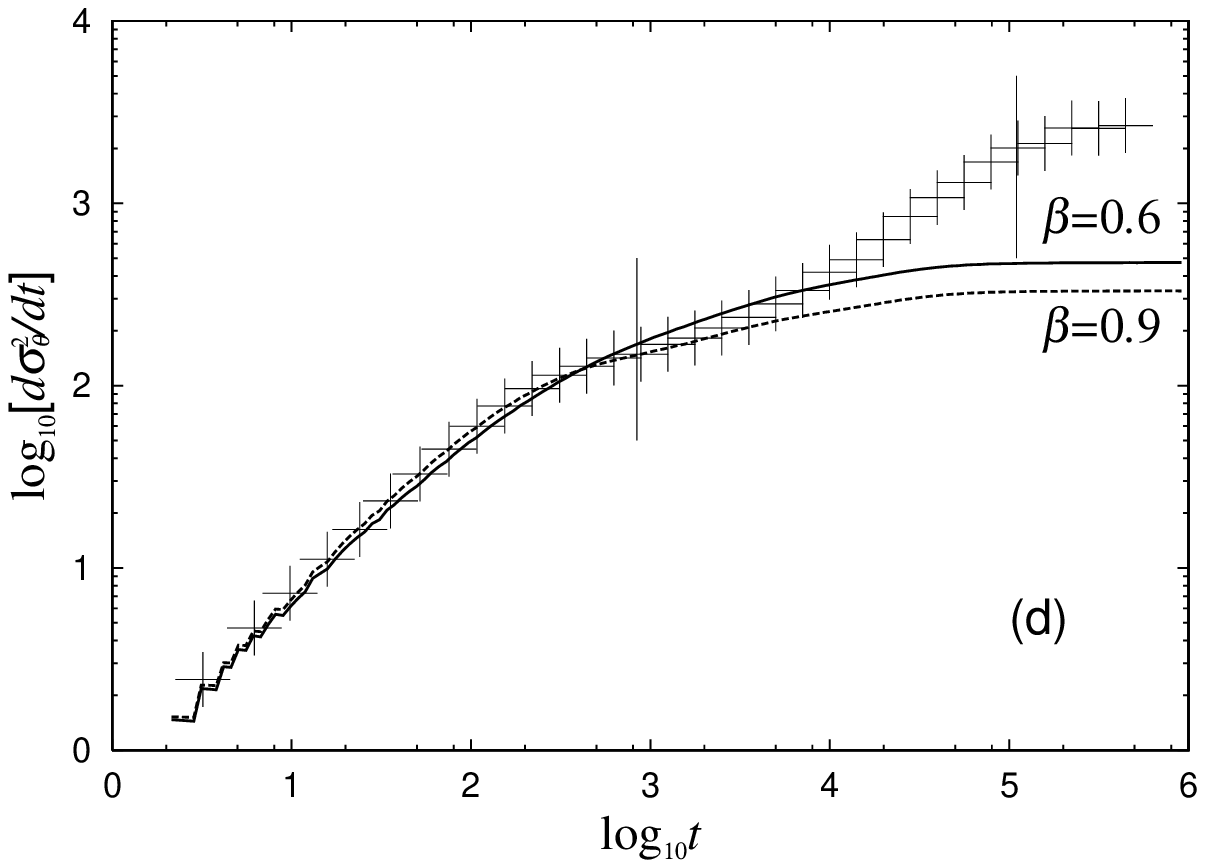}
  \caption{ Time derivative of the mean square displacement ,
    $\rmd\sigma_{\theta}^{2}(t)/\rmd t$.
    (a) Numerical results (crosses) and 
    reproduced one (solid curve) using Eq.(\protect\ref{eq:dsigma-dt}) 
    and the approximate functions of the three parameters 
    in Eqs.(\protect\ref{eq:app-tanh}).
    In (b), (c) and (d), 
    $C_{p}(0;\tau)$, $\tcorr(\tau)$ and $\beta(\tau)$
    are kept constant, respectively.
    In (d), two constants for $\beta(\tau)$ have been tested.
    Solid and dashed curves represent $\beta=0.6$ and $0.9$, respectively.
    The short vertical lines mark the end of Stages-I and II.}
  \label{fig:repro-dsigma}
\end{figure}

The hyperbolic tangent functions are in good agreement with
numerical results, as shown in Fig.\ref{fig:stexp-factor}.
To confirm the validity of the approximation,
we  reproduced $\rmd\sigma_{\theta}^{2}/\rmd t$
using Eqs.(\ref{eq:dsigma-dt}) and (\ref{eq:app-tanh}),
and the reproduced one is in good agreement with numerical results,
as shown in Fig.\ref{fig:repro-dsigma}(a).

If we fix $C_{p}(0;\tau)$ at its middle value, $0.431$,
we find that the dependence on $\tau$ of $C_{p}(0;\tau)$
does not affect significantly $\rmd\sigma_{\theta}^{2}/\rmd t$,
as shown in Fig.\ref{fig:repro-dsigma}(b).
By fixing $\tcorr(\tau)$ at its middle value, $250$,
we obtain the same conclusion for $\tcorr(\tau)$ 
as for $C_{p}(0;\tau)$,
particularly in Stage-II (see Fig.\ref{fig:repro-dsigma}(c)).
On the contrary, if we fix $\beta(\tau)$ at $0.6$ or $0.9$,
we observe no anomaly in diffusion
as shown in Fig.\ref{fig:repro-dsigma}(d),
because $\rmd\sigma_{\theta}^{2}(t)/\rmd t$ is proportional to $t$
and is constant in short and long time regions respectively,
and the same behavior is obtained at equilibrium
(see Fig.\ref{fig:equili}(b)).
Consequently, among the three parameters,
$\beta(\tau)$ plays a crucial role 
to produce anomaly in diffusion.

\subsection{Dependence on Degrees of Freedom}
\label{sec:dep-N}

\begin{figure}[htbp]
  \centering
  \includegraphics[width=7.5cm]{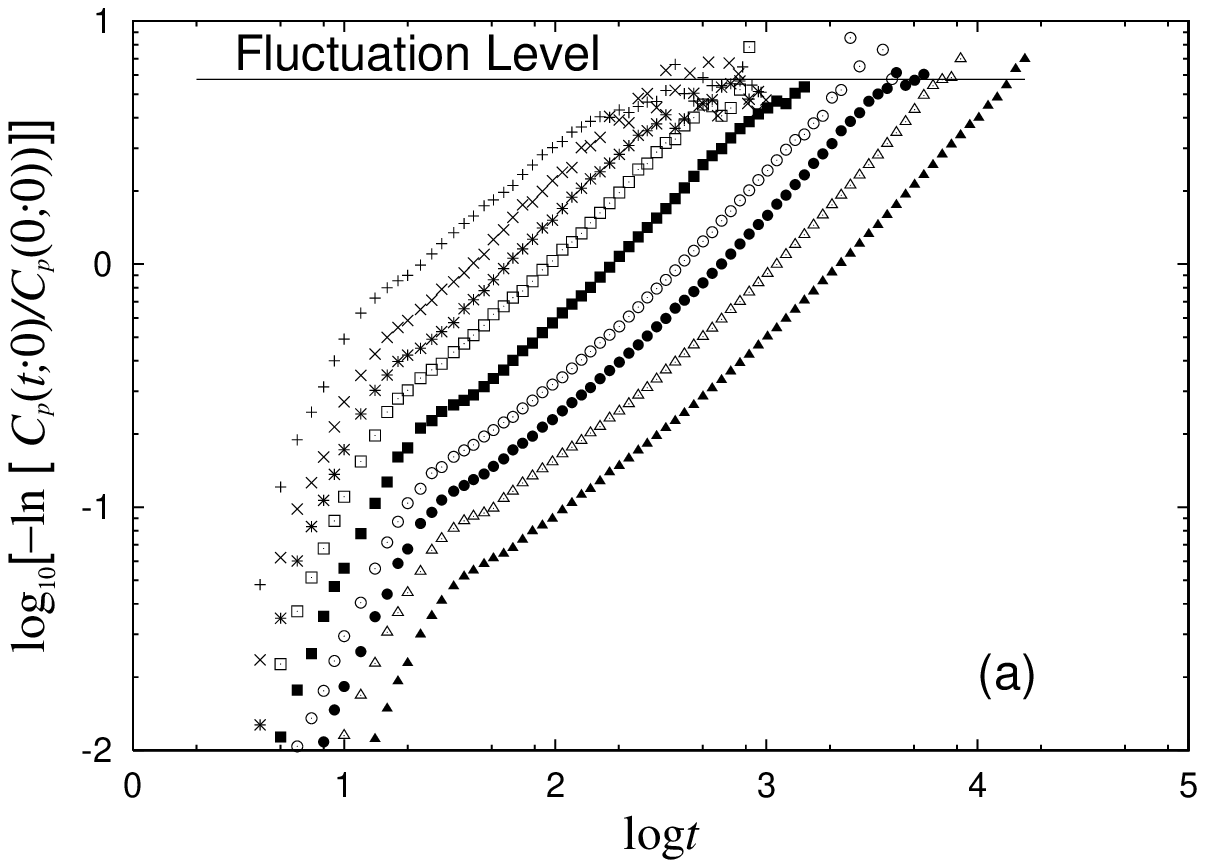}
  \includegraphics[width=7.5cm]{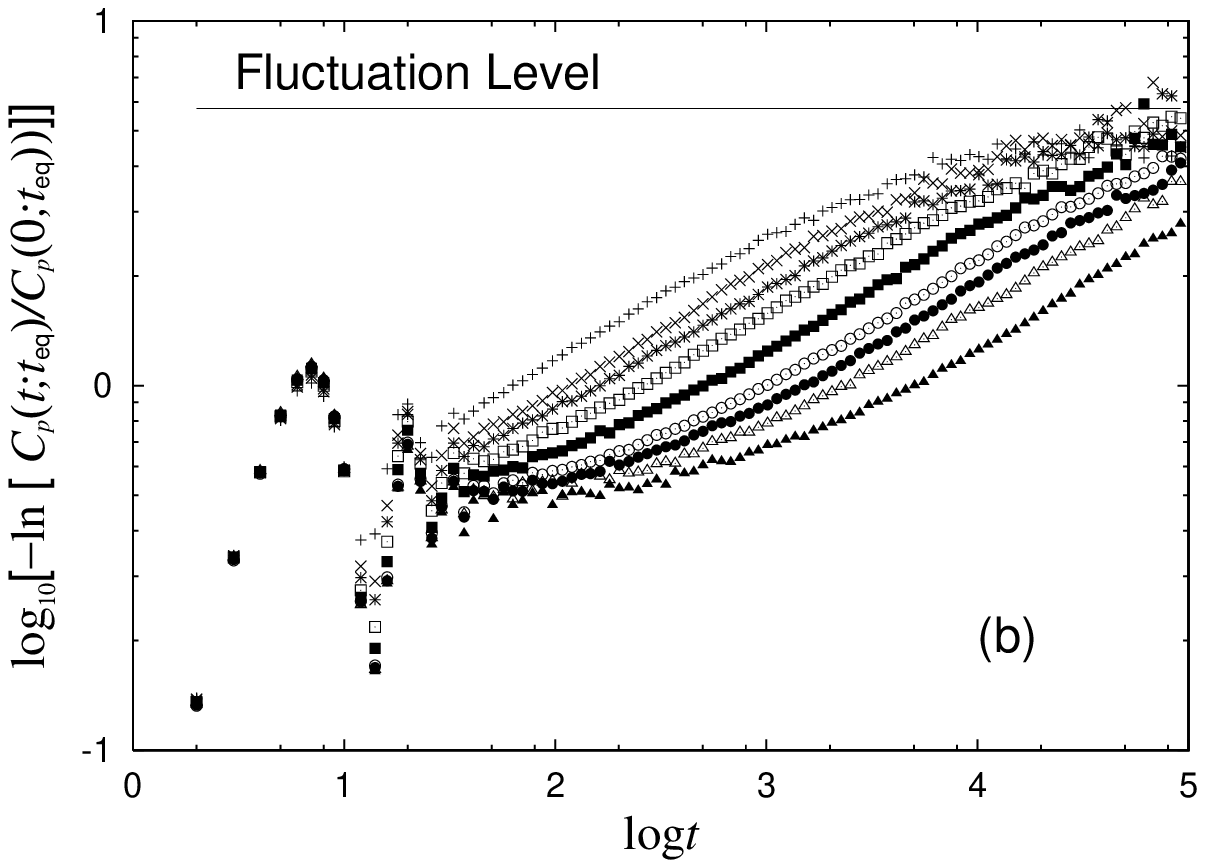}
  \caption{Double log-log plots of correlation functions
    for various values of degrees of freedom $N$.
    (a) $C_{p}(t;0)$ (Stage-I). (b) $C_{p}(t;\teq)$ (Stage-III).
    In both (a) and (b), 
    $N=100(1000)$, $200(500)$, $300(300)$, $500(200)$,
    $1000(100)$, $2000(100)$, $3000(50)$, $5000(10)$ and $10000(10)$ 
    from top to bottom,
    where the inside of parentheses represent numbers of realizations
    for $C_{p}(t;\teq)$. For $C_{p}(t;0)$,
    the number is $1000$ for $N=1000$ and is $100$ for the others.}
  \label{fig:Ndep-Cp}
\end{figure}

\begin{figure}[htbp]
  \centering
  \includegraphics[width=7.5cm]{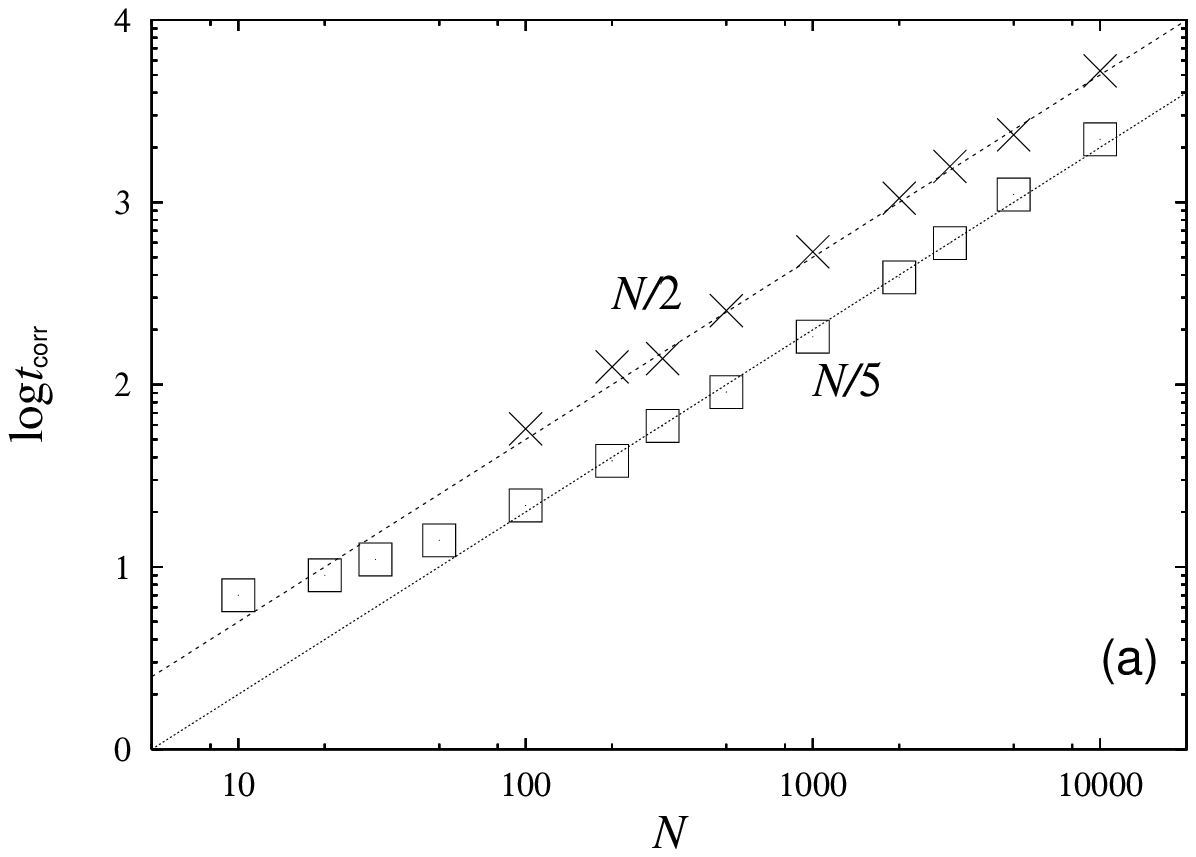}
  \includegraphics[width=7.5cm]{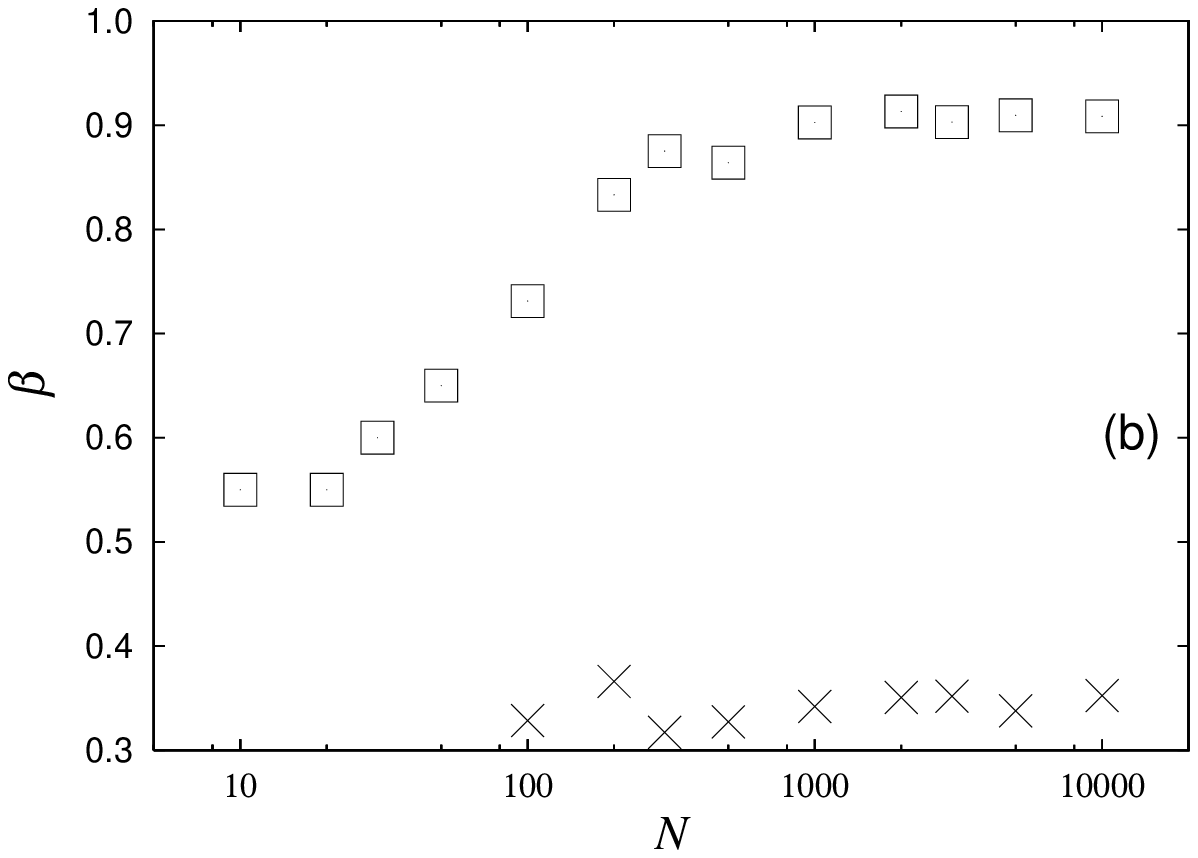}
  \caption{Two parameters of correlation function as functions of 
    degrees of freedom, $\tcorr(N)$ (a) and $\beta(N)$ (b).
    In both (a) and (b), squares ($\boxdot$) represent
    values of the parameters for $C_{p}(t;0)$ (Stage-I),
    and crosses ($\times$) for $C_{p}(t;\teq)$ (Stage-III).}
  \label{fig:Ndep-tcorr-beta}
\end{figure}

In Stage-I (and III), we fit $C_{p}(t;0)$ (resp. $C_{p}(t;\teq)$)
by a stretched exponential function,
which has three parameters: 
$C_{p}(0;0)$, $\tcorr(0)$ and $\beta(0)$
(resp. $C_{p}(0;\teq)$, $\tcorr(\teq)$ and $\beta(\teq)$).
In order to obtain scaling laws for the parameters,
we show them as functions of degrees of freedom $N$.
The parameters $C_{p}(0;0)$ and $C_{p}(0;\teq)$ represent temperature
at $t=0$ and at equilibrium respectively,
and hence they do not depend on $N$.
We therefore focus on the other four parameters,
$\tcorr(0)$, $\beta(0)$, $\tcorr(\teq)$ and $\beta(\teq)$.
The correlation functions, $C_{p}(t;0)$ and $C_{p}(t;\teq)$,
are shown in Fig.\ref{fig:Ndep-Cp},
and values of the four parameters are reported 
as functions of $N$ in Fig.\ref{fig:Ndep-tcorr-beta}.

For large $N$, $N\geq 200$, 
the correlation times are proportional to $N$,
that is $\tcorr(0)=N/5$ and $\tcorr(\teq)=N/2$,
and the stretching exponents $\beta(0)$ and $\beta(\teq)$ 
are almost constants.
We expect that these scaling lows for the four quantities
are kept even in the thermodynamic limit,
although they break for small $N$
where $\tcorr(0)$ is larger than $N/5$ and
$\beta(0)$ is smaller than the constant.
The duration of Stage-I, $t_{\text{I/II}}$,
is around $23$ for $N=100$,
and hence $\tcorr(0)$ and $\beta(0)$ are estimated
mainly not in Stage-I, but in Stage-II from Fig.\ref{fig:Ndep-Cp}(a).
In Stage-II, $\tcorr(\tau)$ and $\beta(\tau)$ 
are increasing and decreasing functions of $\tau$ respectively
(see Fig.\ref{fig:stexp-factor}),
and hence $\tcorr(0)$ and $\beta(0)$ are larger and smaller
than expected values, respectively.

\section{Summary}
\label{sec:summary}

As a summary,
we have investigated the relation between relaxation and diffusion
in a Hamiltonian system with long-range interactions.
The relaxation process is divided into three stages:
quasi-stationary, relaxational and equilibrium.
We showed that diffusion becomes anomalous 
only in the second non-stationary stage,
where magnetization is increasing and goes towards to
the canonical value.
The result mentioned above does not depend on 
the number of degrees of freedom,
at least from $N=100$ to $10000$.

The interval where the anomaly in diffusion appears
becomes shorter and shorter
in logarithmic time scale as $N$ increase 
corresponding to a sharper change of magnetization.
Moreover, a detailed investigation exhibits 
the absence of power-type diffusion even in the non-stationary stage.
We guess that anomaly in diffusion 
is a transient anomaly due to non-stationarity.

Diffusion is obtained by integrating 
the correlation function of momenta  $C_{p}(t;\tau)$
and the correlation function is approximated by a series of
stretched exponential functions 
$C_{p}(t;\tau)=C_{p}(0;\tau)~\exp[-(t/\tcorr(\tau))^{\beta(\tau)}]$.
Among the three parameters, $C_{p}(0;\tau)$, $\tcorr(\tau)$ and $\beta(\tau)$,
the stretching exponent $\beta(\tau)$ plays a crucial role
to yield anomaly in diffusion.
If we assume that $\beta(\tau)$ is a constant,
we never observe anomaly in diffusion.
This result is consistent with the fact that anomaly in diffusion
does not appear in (quasi-)stationary state,
because correlation function $C_{p}(t;\tau)$, 
and $\beta(\tau)$ accordingly, 
are almost invariant with respect to $\tau$.

We also investigated scaling laws concerning degrees of freedom $N$.
The duration of quasi-stationary stage is proportional to $N^{1.7}$,
and relaxation time, at which the system reaches at equilibrium,
is also proportional to $N^{1.7}$ asymptotically
although some corrections must be added.
In both quasi-stationary and equilibrium stages,
$\tcorr$ is proportional to $N$,
and $\beta$ is almost constant. 
These simple scaling laws imply 
that fitting by stretched exponential functions is valid
irrespective of degrees of freedom.

We have not understood the theoretical reason 
of the appearance of a stretched exponential function.
If we assume that several time scales 
with exponential correlation function, $\exp(-t/\tcorr)$, are present,
and we assume probability distribution function of $\tcorr$, $P(\tcorr)$,
then we obtain a stretched exponential function
$\int P(\tcorr) \exp(-t/\tcorr) \rmd\tcorr$
by choosing suitable forms for $P(\tcorr)$ \cite{palmer-84,pelcovits-83}.
In our model, $P(\tcorr)$ corresponds to the distribution of
time scales of individual rotators.
The investigation of the macrovariable $C(t;\tau)$
in relation with the microvariables of the individual particle
correlation functions will be a subject of future work.

\vspace*{1em}
\acknowledgements
I thank Stefano Ruffo for a careful reading of the manuscript
and useful comments.
I acknowledge valuable discussions with
Alessandro Torcini, Freddy Bouchet and Julien Barr\'e.

\end{document}